\documentclass[12pt,a4paper]{jpconf}
\usepackage{iopams,amsmath}
\usepackage[cp1251]{inputenc}            % Выбор языка и кодировки
\usepackage[english]{babel}
\begin{document}
\title{Dynamical equations and transport coefficients for the metals at high pulse electromagnetic fields}

\author{Volkov~N~B$^{1,*}$, Chingina~E~A$^{1}$, and Yalovets~A~P$^{2}$}

\address{$^1$~Institute of Electrophysics, Russian Academy of Science, Ural Branch,\\Amundsen Street 106, Yekaterinburg 620016, Russia}
\address{$^2$~South-Ural State University, Lenin Ave 76, Chelyabinsk 454080, Russia}

\ead{$^{*}$nbv@iep.uran.ru}

\begin{abstract}
We offer a metal model suitable for the description of fast electrophysical processes in conductors under influence of powerful electronic and laser radiation of femto- and picosecond duration, and also high-voltage electromagnetic pulses with picosecond front and duration less than 1 nanoseconds. The obtained dynamic equations for metal in approximation of one quasineutral liquid are in agreement with the equations received by other authors formerly. New wide-range expressions for the electronic conduction in strong electromagnetic fields are obtained and analyzed.
\end{abstract}

\section{Introduction}
Progress in generation of powerful laser pulses of femtosecond and picosecond duration
\cite{Bloembergen:1999}, in creation of experimental high-voltage techniques for generation of electromagnetic pulses (EMP) and intensive electronic beams of picosecond duration
\cite{Mesyats:2005}, and also the results of experimental research of influence on metals of femtosecond laser radiation \cite{Milchberg:1988,Agranat:2007} and the EMP with picosecond front
(see \cite{Barakhvostov:2011} where it is established that the radial electrical intensity $E_{r}$ on the boundary of a metal wire 20~$\mu$m in diameter attains the value $E_{r} = 24$~MV/cm) points to necessity of creation of physical and mathematical models of metal for the description of intensive fast electrophysical processes.

Heating of the conduction electrons takes place under influence of ultrashort intensive laser or electronic radiation on the metal. If the conduction electrons have no time to transfer the acquired energy to the lattice (for example, the time of  electron-phonon energy  exchange  in aluminium according to our estimates equals $\tau_{\varepsilon} \sim 10^{-12} - 10^{-10}$~s (see \cite{Volkov:2001} and references in it)), then their mean free path becomes considerably larger than the average interatomic distance. The fast electrons excite non-equilibrium fluctuations of the lattice  in a metal, i.e. generate the non-equilibrium phonons as a result of  violation of quasi-neutrality. The generation of non-equilibrium phonons leads to effective increase in speed of the electron-phonon exchange of energy and to deceleration of the fast electrons. Such situation is typical for the normal (not superconducting) metals at low temperatures when equilibrium phonons degrees of freedom are frozen i.e. when the number of phonons $N_{p} \rightarrow 0$, and the free-path length of conduction electrons $l(\varepsilon) \rightarrow \infty$ \cite{Brandt:2005}. Therefore, it is necessary to use more complex models for the description of elastic deformations in the metal at low temperatures,  in which the dynamics of metal ions is described by the equations of continuum mechanics, and the dynamics of conduction electrons is described by the kinetic equation\cite{Andreev:1985,Pushkarov:2001}. Intensive ultrashort electron beams excite the elastic strains in a metal due to non-zero weight and non-zero electric charge of particles,  and  lead to destruction of the metal even in the case of weak heating of electrons \cite{Volkov:2007}. It is also necessary to mark that the appearance of the collective quasi-particle excitations like Langmuir plasmons and the ion-sound ones  in classical plasma is one of the main mechanisms of absorption of laser radiation and relaxation of intensive electron beams in plasma \cite{Nezlin:1982,Kingsep:1996} (in a metal, the Langmuir's plasmons are termed just as the plasmons, and the analog of the ion-sound plasmons are phonons \cite{Brandt:2005}).

Thus, the simultaneous existence of slow dynamics of metal ions and fast dynamics of quasi-particle excitations (phonons and conduction electrons) is specific for the interaction of ultrashort pulses of laser and optical radiation, and requires the construction of mesoscopic (single and multiple-speed) models. In these models, the slow movements (lattice deformation) should be described within the framework of multi-liquid and multi-temperature continuum mechanics, and the fast movements should be described by the kinetic equations that take into account the kinetics of mutual transitions of electrons from discrete to continuous states. As a rule, the slow and the fast movements are considered separately in the literature: one-liquid one-temperature models for the slow movements, and kinetic models for the fast movements.

The common methods of constructing the macroscopic mathematical models of one-temperature, one-liquid or single-speed continua, interacting with the gravitational and electromagnetic fields, are considered in detail in \cite{Sedov:1989}. The basic variational equation
\begin{equation}\label{S0}
  \delta \int_{V_{4}} \Lambda dV_{4} + \delta W^{*} + \delta W = 0,
\end{equation}
where $\Lambda$ is the Lagrangian density, which is equal to the total energy density, taken with the minus sign; $\delta W^{*}$ is the  influx of energy to the four-volume which is denoted $V_{4}$; $\delta W$ is the additional inflow of energy due to the power interactions at the surface $\Sigma_{3}$ (the boundary of the volume $V_{4}$).  The equation (\ref{S0}) is actually the energy equation for arbitrary virtual increments of independent parameters of a continuum. General methods of creation of one-temperature multi-speed continua macroscopic models are based on the approach offered by Landau for obtaining of a double-speed hydrodynamic model of superfluid helium-II \cite{Landau:1969}, and are considered in the monograph \cite{Blokhin:1994}. In the work \cite{Andreev:1985} the non-linear theory of elasticity of neutral metal is constructed by means of Landau's method at low temperatures when it is possible to neglect the contribution of phonons to deformation of a crystal. Thus dynamics of the conduction electrons is described in \cite {Andreev:1985} by the kinetic equation. One of us has constructed in \cite{Volkov:1997} the gauge-invariant field model of the current-carrying plasma-like medium (neutral metal) with topological defects (dislocations and disclinations) with the use of the methods developed in \cite{Andreev:1985,Sedov:1989}. In this model the dynamics of conduction electrons and phonons is described in the terms of the kinetic equations.
However, the models described in the works \cite{Andreev:1985,Volkov:1997} are not applicable for the description of fast electrophysical processes in a metal under influence of ultrashort pulses of electromagnetic and electronic radiation, owing to the assumption of conducting medium neutrality, and the neglecting of  phonons contribution to the internal energy of the metal \cite{Andreev:1985}. Also the models \cite{Andreev:1985,Volkov:1997} are not applicable for the description of high-voltage EMP with the picosecond duration for which great values of electrical intensity are characteristic \cite{Barakhvostov:2011}.

Therefore the purpose of present article is creation of mesoscopic two-fluid and two-temperature models in which dynamics of the quasi-particle excitations (phonons and conduction electrons) is described by the kinetic equations. In view of experimentally established fact \cite{Barakhvostov:2011}  the existence of electromagnetic fields with electric field intensity up to 24~MV/cm, in the case of picosecond electromagnetic pulses impact on conductors, we have found as well and analyzed the wide-range expression for the electrical conductivity of metal in a strong electromagnetic field.

\section{Dynamic equations}
In this paper, we will be restricted for simplicity to the case of an ideal (defect-free) non-magnetic crystal. Moreover, when considering the kinetics of mutual transitions of electrons from discrete to continuous state we shall similarly \cite{Derzhiev:1986} neglect of the distinction of the weights of ions with various electric charge z. I.e. we shall believe that a metal consist of two liquids: the ionic liquid with the mean concentration of particles $n=\sum_{z=0}^{Z}n_{z}$, the macroscopical velocity ${\bf v}=\sum_{z=0}^{Z}n_{z}{\bf v}_{z}/n$ and the mean charge $\bar{z}=\sum_{z=0}^{Z}z \cdot n_{z}/n$, and the electronic one with the mean concentration of particles $n_{e}=\langle f_{e}\rangle$ and velocity ${\bf v_{e}}$ (where $f_{e}({\bf p},{\bf r},t)$ and $\bf p$ are the distribution function and the quasimomentum of conduction electrons, respectively; $\langle ... \rangle$ designates averaging on the distribution function). The densities of the electric charge and current are determined by formulas: $\varrho_{e}=e(\bar{z}n-n_{e})$ and ${\bf j}=e(\bar{z}n{\bf v}-n_{e}{\bf v}_{e})$. According to the last ratio, ${\bf v}_{e}$ can by expressed  in ${\bf v}$ and ${\bf j}$ as ${\bf v}_{e}=\bar{z}nn_{e}^{-1} ({\bf v}-{\bf j}(e\bar{z}n)^{-1})$. Then the continuity equations for the ionic and electronic components can be written as:
\begin{equation}\label{Eqn1}
  \frac{\partial n}{\partial t} + \nabla\cdot(n {\bf v}) = 0;
\end{equation}
\begin{equation}\label{Eqn2}
  \frac{\partial n_{e}}{\partial t} + \nabla\cdot(n_{e} {\bf v}) = \Gamma_{e} + \frac{1}{e} \nabla \cdot \left({\bf j} - e \left(\bar{z}n - n_{e}\right){\bf v}\right).
\end{equation}

The equation for the mean charge follows from the continuity equation for the electric charge in view of the equations (\ref {Eqn1}) and (\ref {Eqn2}):
\begin{equation}\label{Eqn3}
  \frac{\partial \bar{z}}{\partial t} + {\bf v}\cdot\nabla\bar{z} = \frac{\Gamma_{e}}{n}.
\end{equation}
In the equations (\ref{Eqn2}) and (\ref{Eqn3}) $\Gamma_{e}=\delta n_{e}/\delta t$ is the rate of the electron density  change as a result of transition from the localized state to the free ("ionization") state and the return transition ("recombination"). In the neutral metal $\bar{z}n=n_{e}$, $\nabla{\bf j}=0$. Therefore in the case of the neutral metal, the equations (\ref{Eqn2}) and (\ref{Eqn3}) are equivalent. In the co-moving frame of reference the stationary equation (\ref{Eqn3}) $\Gamma_{e}=0$ is an analogue of the Saha equation for the dense plasma.

It is known \cite{Andreev:1985} that a crystal lattice is characterized by the translation  vectors ${\bf a}_{\alpha} (\alpha=1, 2, 3)$  and the invariant metric tensors
$g_{\alpha \beta}={\bf a}_{\alpha} {\bf a}_{\beta}$ of the direct lattice and ${\bf a}^{\alpha}~(\alpha=1, 2, 3)$ and $g^{\alpha \beta}={\bf a}^{\alpha} {\bf a}^{\beta}$ of the reciprocal lattice (${\bf a}_{\alpha} {\bf a}^{\beta}=\delta_{\alpha}^{\beta},~{\bf a}_{\alpha i} {\bf a}_{k}^{\alpha}=\delta_{i k}$ (here and below the summation over repeated indices is performed; also ${\bf b} {\bf c}$ and $[{\bf a},{\bf c}]$ are the scalar and cross products of vectors ${\bf b}$ and ${\bf c}$)). Physically infinitesimal, i.e. big in comparison with the period of the lattice, but small in comparison with distance on which the lattice parameters change considerably, the differential of the coordinates $d{\bf r}(t)$ looks like
\begin{equation}\label{Eqn4}
  d{\bf r}={\bf a}_{\alpha}dN^{\alpha}+{\bf v}dt,
\end{equation}
where $N^{\alpha}({\bf r},t)={\bf a}^{\alpha} d{\bf r} - {\bf a}^{\alpha} {\bf v} dt$ are the integer coordinates of the lattice ions  measured in terms of the translation vectors ${\bf a}_{\alpha}$ being one-valued functions ${\bf r}$ only in absence of dislocations in a lattice. From (\ref{Eqn4}) it follows that ${\bf a}^{\alpha}={\bf\nabla}N^{\alpha},
{\bf v}=-{\bf a}_{\alpha}\partial N^{\alpha}/\partial t=-{\bf a}_{\alpha}{\dot{N}}^{\alpha}$.

According to \cite{Andreev:1985} dynamics of the translation vectors is determined by the equations:
\begin{eqnarray}
   {\dot{\bf a}}_{\alpha}&=&-({\bf v} \nabla){\bf a}_{\alpha}+
   ({\bf a}_{\alpha} \nabla){\bf v}, \nonumber \\
   \dot{a}_{i}^{\alpha}&=&-({\bf v} \nabla)a_{i}^{\alpha}-
   a_{k}^{\alpha}\frac{\partial v_{k}}{\partial x_{i}}.
   \label{Eqns5}
\end{eqnarray}

The density of lattice ions $\rho_\ell = M A n$ ($M=1.66\cdot10^{-24}$~g is the atomic mass unit; $A$ is the atomic weight) is connected with the invariant metric tensor by well known relation $\rho_\ell=M A s/V_{c} = M A s/ \sqrt{g}=M A s/\sqrt{\det g_{\alpha\beta}}$, i.e.
$n = s (\det g_{\alpha\beta})^{-1/2}$ (s is the number of atoms in the lattice unit cell). The identity $dg=-g g_{\alpha\beta}dg^{\alpha\beta}$ allows us to be convinced easily that the ion density in this form satisfies to the continuity equation (\ref{Eqn1}).

In quasiclassical approximation one-electron wave function for the conduction electrons is the function of integer coordinates of ions \cite{Andreev:1985,Lifshits:1973}:
\begin{equation}\label{Eqn6}
  \Psi(N^{\alpha},t)\sim\exp\left(\frac{i}{\hbar}S_{0}\left(N^{\alpha},t\right)\right),
\end{equation}
where $S_{0}\left(N^{\alpha},t\right)$ is the classical action. In a periodic motionless lattice the Hamilton function (Hamiltonian) coincides with electron energy $\varepsilon=\varepsilon\left(k_{\alpha},g^{\alpha\beta}\right)$ and it is the periodic function of the invariant quasimomentum $k_{\alpha}$ with the period $2\pi\hbar$:
\begin{equation}\label{Eqn7}
  \frac{\partial S_{0}}{\partial t}=-\varepsilon_{e}\left(k_{\alpha},g^{\alpha\beta}\right), \frac{\partial S_{0}}{\partial N^{\alpha}}=k_{\alpha}.
\end{equation}

The action in a deformable crystal lattice is defined with the help of the Galilean transformation of one-electron wave functions \cite{Landau3:1977}
\begin{equation}\label{Eqn8}
  S=S_{0}+m{\bf v} {\bf r}-\frac{mv^{2}t}{2},
\end{equation}
where $m$ is the conduction electron mass. Its quasimomentum and the Hamiltonian both are defined by differentiation (\ref{Eqn8}) \cite{Andreev:1985}:
\begin{equation}\label{Eqn9}
  {\bf p}=\frac{\partial S}{\partial {\bf r}}|_{t}=k_{\alpha}\nabla N^{\alpha}+
  m{\bf v} \Rightarrow k_{\alpha}={\bf a}_{\alpha} ({\bf p}-m{\bf v});
\end{equation}
\begin{equation}\label{Eqn10}
  H_{e}({\bf p},{\bf r},t)=-\frac{\partial S}{\partial t}|_{{\bf r}}=-k_{\alpha}\dot{N}^{\alpha}+ \varepsilon_{e}+\frac{mv^{2}}{2}=\varepsilon_{e}+{\bf p} {\bf v}-\frac{mv^{2}}{2}.
\end{equation}

In (\ref{Eqn10}) $\varepsilon_{e}=\varepsilon_{e}\left({\bf a}_{\alpha}
\left({\bf p}-m{\bf v}\right),g^{\alpha\beta}\right)$ is a periodic function of the invariant quasimomentum with the period $2\pi\hbar$ in contradistinction to the Hamiltonian. The energy of a conduction electron in a deformable lattice is defined by the Galilean transformation \cite{Andreev:1985}:
\begin{equation}\label{Eqn11}
  \tilde{\varepsilon}_{e}=\varepsilon_{e}+{\bf v} {\bf p}_{0}+
  \frac{mv^{2}}{2}=\varepsilon_{e}+m{\bf v} \frac{\partial\varepsilon_{e}}{\partial {\bf p}}+
  \frac{mv^{2}}{2}.
\end{equation}

Foregoing, the expressions for a conduction electron and the Hamiltonian allow us to write  the kinetic equation with excluded acyclic on invariant quasimomentum $k_{\alpha} $ variables as \cite{Andreev:1985}:
\begin{equation}\label{Eqn12}
 \frac{\partial f_{e}}{\partial t}+\left({\bf v} \nabla\right)f_{e}+
 {\bf a}_{\alpha} \nabla f_{e}\frac{\partial\varepsilon_{e}}{\partial k_{\alpha}}-
 \frac{\partial f_{e}}{\partial k_{\alpha}}\left({\bf a}_{\alpha}
 \frac{\partial \varepsilon_{e}}{\partial{\bf r}}+e{\bf a}_{\alpha}
 {\bf E}'+\frac{e}{c}\frac{\partial \varepsilon_{e}}{\partial k_{\beta}}
  {\bf H}' \left[{\bf a}_{\alpha},{\bf a}_{\beta}\right]\right)=\hat{I}f_{e},
\end{equation}
where $\hat{I}f_{e}$ is the integral operator of elastic and unelastic electron
collisions;
$${\bf H}'={\bf H}-\frac{mc}{e}[\nabla,{\bf v}],~{\bf E}'={\bf E}+\frac{1}{c}[{\bf v},{\bf H}']+
\frac{m}{e}\left(\frac{\partial {\bf v}}{\partial t}+\frac{\partial}{\partial {\bf r}} \left(\frac{v^{2}}{2}\right)\right).$$

The Hamiltonian for the phonon gas in deformable metal can be represented as the function of the invariant phonon quasimomentum $q_{\alpha}={\bf a}_{\alpha} {\bf p}~(\alpha=1, 2, 3)$:
\begin{equation}\label{Eqn13}
  H_{p}=\varepsilon_{p}\left({\bf a}^{\alpha}q_{\alpha},g^{\alpha\beta}\right)+
  q_{\alpha}{\bf a}^{\alpha} {\bf v}=
  \hbar\omega\left({\bf a}^{\alpha}q_{\alpha},g^{\alpha\beta}\right)+
  q_{\alpha}{\bf a}^{\alpha} {\bf v}.
\end{equation}
Taking into account (\ref{Eqn13}) we can write the kinetic equation for phonon gas  similar to the equation (\ref{Eqn12}) as:
\begin{equation}\label{Eqn14}
  \frac{\partial f_{p}}{\partial t}+\left({\bf v} \nabla\right)f_{p}+
 {\bf a}_{\alpha} \nabla f_{p}\frac{\partial\varepsilon_{p}}{\partial q_{\alpha}}-
 \frac{\partial f_{p}}{\partial q_{\alpha}}{\bf a}_{\alpha}
 \frac{\partial \varepsilon_{p}}{\partial{\bf r}}=\hat{I}f_{p},
\end{equation}
where $\hat{I}f_{p}$ is the integral operator of the phonon collisions among themselves and with the conduction electrons.

The mean value of the phonon gas energy is determined by the relation \cite{Landau5:1980}:
\begin{equation}\label{Eqn15}
  \langle\varepsilon_{p}\rangle=\sum_{\beta}\int\hbar\omega_{\beta}f_{p}\frac{d^{3}p}{(2\pi\hbar)^{3}}=
  \int\hbar\omega f_{p}g_{p}(\omega)d\omega,
\end{equation}
where $\beta$ is the number of a normal oscillation branch of the crystal lattice; $g_{p}(\omega)$ is its phonon spectrum determined by the structure of the crystal lattice and satisfies an integral condition:
\begin{equation}\label{Eqn16}
  \int g_{p}(\omega)d\omega = 3 N s.
\end{equation}
In the expression (\ref{Eqn16}) $N$ is the general number of the lattice unit cells.

Let us determine the macroscopical momentums and the energies of the lattice (ionic component)
(${\bf P}^{(\ell)}$ and $W^{(\ell)}$) and the conduction electrons (${\bf P}^{(e)}$ and $W^{(e)}$) by the following expressions:
\begin{eqnarray}\label{Eqn16-1}
{\bf P}^{(\ell)}=MAn{\bf v};\quad {\bf P}^{(e)}=m\langle f_{e}\rangle
\left({\bf v}+\frac{1}{\langle f_{e}\rangle}
\Big \langle\frac{\partial \varepsilon_{e}}{\partial {\bf p}}f_{e}\Bigr \rangle\right)=
m\langle f_{e} \rangle {\bf v}_{e};\nonumber\\
W^{(\ell)}=\frac{MAnv^{2}}{2}+\varepsilon_{\ell}\left(g^{\alpha\beta}\right)+
\bigl \langle \varepsilon_{p}f_{p} \bigr \rangle;\quad W^{(e)}=\bigl \langle \tilde{\varepsilon}_{e}f_{e}\bigr\rangle=
\frac{mv^{2}}{2}+m{\bf v}
\Bigl\langle\frac{\partial \varepsilon_{e}}{\partial {\bf p}}f_{e}\Bigr\rangle+
\bigl\langle \varepsilon_{e}f_{e}\bigr\rangle,
\end{eqnarray}
where $\varepsilon_{\ell}\left(g^{\alpha\beta}\right)$ is the elastic ("cold") energy of the lattice.

Being based on the approaches advanced in \cite{Klimontovich:1986,Braginskii:1965}, the dynamic equations for a metal in two-liquid approximation can be written as:
\begin{equation}\label{Eqn17}
  \frac{\partial {\bf P}^{(\ell)}}{\partial t}+\nabla {\bf \Pi}^{(\ell)}=e\bar{z}n
  \biggl({\bf E}+\frac{1}{c}[{\bf v},{\bf H}]\biggr)+{\bf R};
\end{equation}
\begin{equation}\label{Eqn18}
  \frac{\partial {\bf P}^{(e)}}{\partial t}+\nabla {\bf \Pi}^{(e)}=-e\bar{z}n
  \biggl({\bf E}+\frac{1}{c}[{\bf v},{\bf H}]\biggr)-{\bf R};
\end{equation}
\begin{equation}\label{Eqn19}
  \frac{\partial W^{(\ell)}}{\partial t}+\nabla {\bf Q}^{(\ell)}=e\bar{z}n\left({\bf E}\cdot
  {\bf v}\right)+Q_{\triangle T};
\end{equation}
\begin{equation}\label{Eqn20}
  \frac{\partial W^{(e)}}{\partial t}+\nabla {\bf Q}^{(e)}=-e\bar{z}n\left({\bf E}
  {\bf v}\right)+{\bf R} \left({\bf v}_{e}-{\bf v}\right)-Q_{\triangle T}+Q_{ne}-Q_{rad},
\end{equation}
where
\begin{equation}\label{Eqn21}
  {\bf R}=\bigl\langle{\bf p}\hat{I}f_{e}\bigr\rangle=m\nu_{ep}\bigl\langle f_{e}\bigr\rangle\left({\bf v}_{e}-{\bf v}\right)=m\nu_{ep}
  \biggl \langle\frac{\partial \varepsilon_{e}}{\partial {\bf p}}f_{e}\biggr\rangle=m\nu_{ep}  \left(\bar{z}n-\bigl\langle f_{e}\bigr\rangle\right){\bf v}-
  \frac{m\nu_{ep}}{e}{\bf j}
\end{equation}
is the change of the ionic component momentum as the result of interaction with the conduction electrons; $\nu_{ep}$ is the effective frequency of the conduction electrons dispersion on the density fluctuations (generally it can depend on the electric field intensity);
${\bf \Pi}^{(\ell)},\quad {\bf \Pi}^{(e)}$ and ${\bf Q}^{(\ell)},\quad {\bf Q}^{(e)}$  are the densities of momentum and energy flows for  ionic and electronic components, accordingly. The last four terms in the equation (\ref{Eqn20}) take into account the change of the conduction electrons energy as the result of their elastic and unelastic collisions:
\begin{equation}\label{Eqn22}
  \bigl\langle\tilde{\varepsilon}_{e}\hat{I}f_{e}\bigr\rangle={\bf R} \left({\bf v}_{e}-{\bf v}\right)-Q_{\triangle T}+Q_{ne}-Q_{rad}=\frac{m\nu_{ep}}{\bigl\langle f_{e}\bigr\rangle}
  \Bigl\langle\frac{\partial \varepsilon_{e}}{\partial {\bf p}}f_{e}\Bigr\rangle^{2}-
  \bigl\langle\varepsilon_{p}\hat{I}f_{p}\bigr\rangle+Q_{ne}-Q_{rad}.
\end{equation}
In (\ref{Eqn22}) the first term is the Joule energy source (it is clearly visible in the case of neutral metal wherein $\bigl<f_{e}\bigl(\partial \varepsilon_{e}/\partial {\bf p}\bigr)\bigr>=-{\bf j}/e$); $Q_{\Delta T}=\bigl<\varepsilon_{p}\hat{I}f_{p}\bigr>=\delta
\bigl<\varepsilon_{e}\hat{I}f_{e}\bigr>$ is the heat exchange between electronic and ionic components ($\delta$ is the efficiency of electron-ion heat exchange); $Q_{ne}$ is the energy change as the result of interband transitions (ionization-recombination processes); $Q_{rad}$ is the radiation energy losses as the result of the conductivity electrons deceleration.

Operating similarly to authors of the article \cite{Andreev:1985} we shall obtain the expressions for densities of momentum and energy flows:
\begin{equation}\label{Eqn23}
  \Pi_{ik}^{(\ell)}=MAnv_{i}v_{k}+\sigma_{ik}^{(\ell)}, \quad \sigma_{ik}^{(\ell)}=
  2\left(\sigma_{\alpha\beta}^{(\ell)}+
  \bigl\langle\lambda_{\alpha\beta}^{(p)}f_{p}\bigr\rangle\right)a_{i}^{\alpha}a_{k}^{\beta}-
  \varepsilon_{\ell}\delta_{ik};
\end{equation}
\begin{eqnarray}\label{Eqn24}
  \Pi_{ik}^{(e)}=-T_{ik}^{(em)}+m\bigl\langle f_{e}\bigr\rangle v_{i}v_{k}+
  m\left(\bar{z}n-\bigl\langle f_{e}\bigr\rangle\right)\left(v_{i}v_{k}-v_{k}v_{i}\right)-
  \frac{m}{e}\left(v_{i}j_{k}+v_{k}j_{i}\right)+\sigma_{ik}^{(e)},\\\nonumber
  \sigma_{ik}^{(e)}=2\bigl\langle\lambda_{\alpha\beta}^{(e)}f_{e}\bigr\rangle
  a_{i}^{\alpha}a_{k}^{\beta};
\end{eqnarray}
\begin{equation}\label{Eqn25}
  {\bf Q}^{(\ell)}={\bf v}\left(\varepsilon_{\ell}+\bigl\langle\varepsilon_{p}f_{p}\bigr\rangle\right)-
  \frac{v^{2}}{2}{\bf P}^{(\ell)}+{\bf v} {\bf \Pi}^{(\ell)}+
  \Bigl\langle\varepsilon_{p}\frac{\partial \varepsilon_{p}}{\partial {\bf q}}f_{p}\Bigr\rangle;
\end{equation}
\begin{equation}\label{Eqn26}
 {\bf Q}^{(e)}={\bf v}\bigl\langle\varepsilon_{e}f_{e}\bigr\rangle -
  \frac{v^{2}}{2}{\bf P}^{(e)}+{\bf v} \bigl({\bf \Pi}^{(e)}+{\bf T}^{(em)}\bigr)+
  \Bigl\langle\varepsilon_{e}\frac{\partial \varepsilon_{e}}{\partial {\bf p}}f_{e}\Bigr\rangle,
\end{equation}
where $\sigma_{\alpha\beta}^{(\ell)}=\partial \varepsilon_{\ell}/\partial g^{\alpha\beta}$;
$\lambda_{\alpha\beta}^{(e)}=\partial \varepsilon_{e}/\partial g^{\alpha\beta}$;
$\lambda_{\alpha\beta}^{(p)}=\partial \varepsilon_{p}/\partial g^{\alpha\beta}$;
$T_{ik}^{(em)}=\bigl(E_{i}E_{k}+H_{i}H_{k}+\delta_{ik}(E^{2}+H^{2})/2\bigr)/(4\pi)$ is
the Maxwell tension tensor \cite{Landau2:1980}.

From two equations of dynamics (\ref{Eqn17})-(\ref{Eqn18}) it is possible to proceed to another two equations, one of which describes dynamics of the metal in the center-of-mass system for ions - conduction electrons, and the second represents the generalized Ohm law where the electron inertia is taken into account:
\begin{equation}\label{Eqn27}
  \frac{\partial {\bf P}^{(\Sigma)}}{\partial t}+\nabla {\bf \Pi}^{(\Sigma)}=0;
\end{equation}
\begin{equation}\label{Eqn28}
  \frac{\partial {\bf j}}{\partial t}=\frac{e^{2}\bar{z}n}{m}\left({\bf E}+
  \frac{1}{c}[{\bf v},{\bf H}]\right)-\frac{e}{mc}[{\bf j},{\bf H}]+e\bigl(\bar{z}n-\langle f_{e}\rangle\bigr)\nu_{ep}{\bf v}+
  en{\bf v}\left(\frac{\Gamma_{e}}{n}-{\bf v}\cdot\nabla\bar{z}\right)-
  \frac{e}{m}{\bf \nabla} {\bf \Pi}^{(e)}-\nu_{ep}{\bf j},
\end{equation}
where ${\bf P}^{(\Sigma)}={\bf P}^{(\ell)}+{\bf P}^{(e)}$,
${\bf \Pi}^{(\Sigma)}={\bf \Pi}^{(\ell)}+{\bf \Pi}^{(e)}$.
At development of the equation (\ref{Eqn28}) we assumed that  $m/(MA)\ll1$. The second term in (\ref{Eqn28}) takes into account the contribution to a current density of the Hall effect, and the next-to-last is the thermoelectricity.

The equations (\ref{Eqn17})-(\ref{Eqn20}) or (\ref{Eqn27}), (\ref{Eqn28}), (\ref{Eqn19}), (\ref{Eqn20}) with the defining ratios (\ref{Eqn16-1}), (\ref{Eqn23})-(\ref{Eqn26}) are the main content of the two-fluid, two- temperature model of the metal offered by us for the description of the fast electrophysical processes. For  receiving those equations we did not do any assumptions of the magnitude of the electromagnetic field, and about of the speed of its change. It is obvious that macroscopic velocities of the electronic and ionic components of the metal have to be essentially less than the light speed. Also the received equations should be supplemented with the complete set of Maxwell's equations \cite{Landau2:1980,Landau8:1984}, and with the initial and boundary conditions (different for each specific objective).

Expressions (\ref{Eqn23})-(\ref{Eqn26}) show that to close the equations set
(\ref{Eqn17})-(\ref{Eqn20}) or (\ref{Eqn27})-(\ref{Eqn28}) and (\ref{Eqn19})-(\ref{Eqn20}) it is necessary to solve the kinetic equations (\ref{Eqn12}) and (\ref{Eqn14}). That will be done in the following section.

\section{The electronic conduction in high pulse electromagnetic fields}
Let us consider the electron transport in metal plasma in a strong electromagnetic field. For this purpose we shall take advantage of the kinetic equation for conduction electrons (\ref{Eqn12}), written down for of the conduction electron distribution function $f_{e}=f_{0}+\delta f=f_{0}+{\bf p} {\bf f_{1}}/|{\bf p}|$ in the co-moving frame of reference in the diffusion approximation \cite{Shabanskii:1954,Shabansky:1957,Ablekov:1980}:
\begin{equation}\label{Eqn32}
  \frac{\partial f_{0}}{\partial t}+\frac{v_{e}}{3}{\bf \nabla} {\bf f}_{1}+
  \frac{1}{n_{e}(\varepsilon_{e})}
  \left(n_{e}(\varepsilon_{e}) \Biggl(\frac{e v_{e}}{3}\bigl({\bf E} {\bf f}_{1}\bigr)-\frac{2mv_{s}^{2}\varepsilon_{e}v_{e}}{l(\varepsilon_{e})}
  \biggl(\frac{f_{0}\bigl(1-f_{0}\bigr)}{T}+
  \frac{\partial f_{0}}{\partial \varepsilon_{e}}\biggr)\Biggr)\right)=0;
\end{equation}
\begin{equation}\label{Eqn33}
  \frac{\partial {\bf f}_{1}}{\partial t}+v_{e}{\bf \nabla}f_{0}+ev_{e}
  {\bf E}\frac{\partial f_{0}}{\partial \varepsilon_{e}}+
  \frac{e}{mc}[{\bf H},{\bf f}_{1}]+\frac{v_{e}}{l(\varepsilon_{e})}{\bf f}_{1}=0,
\end{equation}
where $v_{s}$ is the sound speed in the metal, determined by the structural part of pressure (without taking into account the pressure of the electron thermal excitation); $n(\varepsilon_{e})=4\pi p^{2}\partial p/\partial \varepsilon_{e}=4\pi m^{3/2} (2\varepsilon_{e})^{1/2}$ is the density of electron states for a parabolic power spectrum  $\varepsilon_{e}=p^{2}/(2m)$; $l(\varepsilon_{e})$ is the length of the conduction electron free path. The equations (\ref{Eqn32}) and (\ref{Eqn33}) are valid under the condition $\delta\ll 1$ ($\delta=m v_{s}^{2}/T$ is the value of transmitted energy from the electrons to the lattice being true at  $T\gg T_{D}$ \cite{Ginzburg:1955}, where $T_{D}$ is the Debye temperature). Since in the ideal gas (plasma) $v_{s}^{2} \sim T/(MA)$ then also in this case $\delta \sim m/(MA) \ll 1$. An additional condition of applicability of the equations (\ref{Eqn32}) and (\ref{Eqn33}) coordinated with the smallness of $\delta$  is the smallness of ratio  $f_{1}/f_{0}$. Its maximal value $(f_{1}/f_{0})_{max} \sim \delta$ \cite{Ablekov:1980} is achieved at $|{\bf E}|\sim mv_{s}^{2}/\bigl(e l(\varepsilon_{e})\bigr)$. The equations (\ref {Eqn32}) and (\ref {Eqn33}) are correct for any electrical field strengths when these two conditions are satisfied. According to Ginzburg~V~L \cite{Ginzburg:1970} the nonlinearity contribution to processes of electronic transport in electromagnetic fields becomes essential at $E \gg E_{p} = \left(3 m T e^{-2} \delta (\omega^{2} + \nu_{ep0}^{2}) \right)^{1/2} \geq E_{p0} = \nu_{ep0}\sqrt{3mT\delta}/e$ (where $\omega,\quad\nu_{ep0}$ are the frequencies of the field and of the electron-ion's collisions at $E = 0$,
respectively). For $\delta = m v_{s}^{2}/T$ \cite{Ginzburg:1955} $E_{p0} = \sqrt{3} m v_{s} \nu_{ep0} / e$. The estimates of $E_{p0}$  for aluminum and copper under normal conditions are values: $E_{p0}(Al) = 5.172$~kV/cm, $E_{p0}(Cu) = 3.842$~kV/cm. The top estimation of a plasma electric field value $E_{pF} = m v_{F} \nu_{ep0} / e$, at which degeneration of the electronic component of metal vanishes, is the following: $E_{pF}(Cu) = 0.893$~MV/cm, $E_{pF}(Al) = 1.149$~MV/cm. Comparing these estimates to the experimental estimation of the radial electric field strength from the work \cite{Barakhvostov:2011} it is possible to state that electronic transfer  will non-linearly depend on the electric intensity, at least, in the surface layer of the metal subjected to influence of a picosecond high-voltage electromagnetic pulse.

Taking into account the aforesaid we shall be limited for simplicity to the case of an uncharged metal of cubic symmetry and homogeneous electromagnetic non-relativistic field with characteristic frequencies $\nu \ll \nu_{ep0}$, when finding-out of influence of a strong electromagnetic field on the electric conduction. In this case the temporal and spatial derivatives and also the term containing $\bf H$ should be neglected. Then the system of the equations (\ref{Eqn32}) and (\ref{Eqn33}) will become:
\begin{equation}\label{Eqn34}
  \frac{e v_{e}}{3}\bigl({\bf E} {\bf f}_{1}\bigr)-\frac{2mv_{s}^{2}\varepsilon_{e}v_{e}}{l(\varepsilon_{e})}
  \biggl(\frac{f_{0}\bigl(1-f_{0}\bigr)}{T}+
  \frac{\partial f_{0}}{\partial \varepsilon_{e}}\biggr)=0;
\end{equation}
\begin{equation}\label{Eqn35}
  ev_{e}{\bf E}\frac{\partial f_{0}}{\partial \varepsilon_{e}}+
  \frac{v_{e}}{l(\varepsilon_{e})}{\bf f}_{1}=0.
\end{equation}
Its solution looks like:
\begin{equation}\label{Eqn36}
  f_{0}(\varepsilon_{e})=\Biggl(exp\biggl(
  \int_{\mu}^{\varepsilon_{e}}\frac{d\epsilon}{T(\epsilon)}\biggr)+1\Biggr)^{-1};~
  ~{\bf f}_{1}(\varepsilon_{e})=-e l(\varepsilon_{e}){\bf E}
  \frac{f_{0}(1-f_{0})}{T(\varepsilon_{e})},
\end{equation}
where
\begin{equation}\label{Eqn37}
  T(\varepsilon_{e}) = T\left(1+
  \frac{e^{2} l^{2}(\varepsilon_{e}) E^{2}}{6 m v_{s}^{2} \varepsilon_{e}}\right)
  = T\left(1 + \frac{e^{2} \tau^{2}(\varepsilon_{e}) E^{2}}{3 m^{2} v_{s}^{2}}\right).
\end{equation}

Using well known definition of the electric current density for a cubic lattice, we shall receive the general expression for the electric conduction:
\begin{eqnarray}\label{Eqn38}
  \sigma(E^{2}) & = & \frac{(2m)^{1/2}e^{2}}{3 \pi^{2} \hbar^{3}}
  \int_{0}^{\infty}   \frac{\varepsilon_{e}^{3/2} \tau(\varepsilon_{e})f_{0}(1-f_{0})d\varepsilon_{e}}  {T\biggl(1+\frac{e^{2}\tau^{2}(\varepsilon_{e})E^{2}}{3m^{2}v_{s}^{2}}\biggr)}\nonumber\\
  & & = \frac{e^{2}\bar{z}n}{3m}\frac{\int_{0}^{\infty}
  \frac{\varepsilon_{e}^{3/2} \tau(\varepsilon_{e})f_{0}(1-f_{0})d\varepsilon_{e}}  {T\biggl(1+\frac{e^{2}\tau^{2}(\varepsilon_{e})E^{2}}{3m^{2}v_{s}^{2}}\biggr)}}
  {\int_{0}^{\infty}\varepsilon_{e}^{1/2}f_{0}d\varepsilon_{e}}=
  \frac{e^{2}\bar{z}n\tau_{ep}}{m}.
\end{eqnarray}
The mean time of the conduction electron dispersion follows from the equation (\ref{Eqn38}):
\begin{equation}\label{Eqn39}
  \tau_{ep}(E^{2})=\frac{1}{3}\frac{\int_{0}^{\infty}
  \frac{\varepsilon_{e}^{3/2} \tau(\varepsilon_{e})f_{0}(1-f_{0})d\varepsilon_{e}}  {T\biggl(1+\frac{e^{2}\tau^{2}(\varepsilon_{e})E^{2}}{3m^{2}v_{s}^{2}}\biggr)}}{\int_{0}^{\infty}\varepsilon_{e}^{1/2}f_{0}d\varepsilon_{e}}.
\end{equation}

The equations (\ref{Eqn36})-(\ref{Eqn38}) show that it is necessary to define the time of electron dispersion $\tau(\varepsilon_{e})$ for finding the electron distribution function and electric conductivity in explicit form. In this work,  we consider collective quasi-particle excitations in the condensed substance as conduction electrons. As shown in the works \cite{Ginzburg:1955a,Silin:1956,Silin:1959} it is possible to consider the conduction electrons as an ideal Fermi gas, i.e. to ignore the Fermi-liquid effects at calculation of the electronic coefficients. In this case, for finding of the free path of conduction electrons we can use analogy, between scattering of the X-ray radiation and conduction electrons on density fluctuations. This analogy was first introduced and used by
J~I~Frenkel \cite{Frenkel:1932} for the explanation of temperature dependence of the specific resistance of metals. This analogy also is neither more nor less than an essence of so-called diffraction model of a metal \cite{Harrison:1966} and of the Ziman's formula \cite{Ziman:1972,Ziman:1979}. For the first time this analogy was used by one of us in 1975-1977 for finding wide-range expression for the electrical conductivity within the plasma model of metals \cite{Volkov:1979} and was applied to the interpretation of experiments on the magnetic cumulation \cite{Volkov:1982}, and also was used for modeling of the magnetohydrodynamic regime of the electrode evaporation in the plasma focus \cite{Gerusov:1982}. Later, the wide-range expressions for the  electronic transport coefficients were offered by Lee and More \cite{Lee:1984} and by Bespalov and Polishchuk (a model offered by them was described in detail in the collective monograph \cite{Ebeling:1991}). An essential difference of our approach is that the static structure factor in a long-wave limit coincided with a known definition of density fluctuations via the isothermal module of compressibility, i.e. it was coordinated with an equation of state. It should be noted also that now-days the methods {\it ab inition} of numerical calculation of the electronic  transport and optical properties of the dense plasma, within the theory of the linear response by means of quantum-mechanical Kubo-Greenwood's formulas (see for example \cite{Desjarlais:2002,Mazevet:2005,Knyazev:2014}),  which practically do not demand the preliminary use of an experiment, are widely developed. Despite the reached results, this method is not applicable in the case of intensive electromagnetic fields owing to absence of the theory of the nonlinear response. Besides, that this method demands large computing expenses for obtaining the reliable results as it is necessary to consider  ensembles with large number of atoms.  On the contrary, the kinetic equations in principle are deprived of these problems and can be used for the solution of strongly nonlinear problems (see for example \cite{Silin:1998}).

Earlier the transport phenomena in high static and time-dependent electric fields were intensively investigated for the semiconductors \cite{Davydov:1937,Bass:1975,Dykman:1981} and the fully ionized classical plasma (see for example \cite{Keishiro:1967,Salat:1969}). The electrical conductivity of metals in strong electric fields at low temperatures has been investigated by Shabanskii~V~P \cite{Shabansky:1954,Shabansky:1957}. The negative result of our search in the Internet has shown that up to now there are no researches of the electrical conductivity and other transport coefficients dependence upon the electric-field strength  in the wide range of pressures and temperatures including the transition region from the metal state to the classical plasma.

 Therefore we shall take advantage of results of one of us  \cite{Volkov:1991,Volkov:2006} in which within the framework of the plasma model the wide-range expression for time of the conduction electrons dispersion on the density  fluctuations having correct asymptotics for the metal and classical plasma is determined as:
\begin{equation}\label{Eqn40}
  \tau(\varepsilon_{e})=\frac{m^{1/2}\varepsilon_{e}^{3/2}G}{2^{1/2}\pi n e^{4}\bar{z}^{2}\Lambda_{eff}},
\end{equation}
where $G=\kappa_{s}/(nT)=MAn\left(\partial P_{s}/\partial \rho\right)|_{T}/(nT)=MAv_{s}^2/T$ (for $T \rightarrow 0~G\propto T^{-1}$, and for $T \rightarrow \infty~G \rightarrow 1$) is a long-wave structure factor  determining by density fluctuations. For finding the expression (\ref{Eqn40}) an effective potential was also used in which dispersion on ionic skeleton electrons was taken into account. The Coulomb logarithm analogue included in (\ref{Eqn40}) is expressed by ratios:
\begin{equation}\label{Eqn41}
  \Lambda_{eff}=\Lambda_{1}-2(Z-\bar{z})\bar{z}^{-1}\Lambda_{2}+
  ((Z-\bar{z})/\bar{z})^{2}\Lambda_{3},
\end{equation}
where
\begin{eqnarray}\label{Eqn42}
  \Lambda_{1}(4k^{2}k_{D}^{-2})=\frac{1}{2}\left(\ln\biggl(1+\frac{4k^{2}}{k_{D}^{2}}\biggr)-
  \frac{4k^{2}k_{D}^{-2}}{1+4k^{2}k_{D}^{-2}}\right);\\ \nonumber
  \Lambda_{2}=\frac{1}{2}\ln(1+4k^{2}r_{cd}^{2})\left(\frac{k_{D}^{2}r_{cd}^{2}}{k_{D}^{2}r_{cd}^{2}-1}
  \ln\frac{1+4k^{2}k_{D}^{-2}}{1+4k^{2}{r_{cd}^{2}}}-
  \frac{1}{k_{D}^{2}r_{cd}^{2}-1}\right); \\ \nonumber
  \Lambda_{3}=\Lambda_{1}(4k^{2}r_{cd}^{2});
\end{eqnarray}
$r_{cd}=r_{c}/(1+k_{D}r_{c})$; $k_{D}^{2}=\min\{k_{s}^{2},k_{ei}^{2}\},~k_{ei}^{2}=k_{De}^{2}+
(k_{Di}^{4}+k_{s}^{4})k_{Di}^{-2};~k_{s}^{2}=r_{s}^{-2}=\bigl(3/(4\pi n)\bigr)^{-2/3};~
k_{Di}^{2}=4\pi e^{2}\bar{z}^{2}n/T;~k_{De}^{2}=2\pi e^{2}\bar{z}nT^{-1}I_{-1/2}(\mu/T)/I_{1/2}(\mu/T);~
I_{\nu}(y)=\int_{0}^{\infty}\bigl(\exp(\xi-y)+1\bigr)^{-1}\xi^{\nu}d\xi;~y=\frac{\mu}{T}$.

Then with the account of (\ref{Eqn40}), the expression (\ref{Eqn37}) for $T(\varepsilon_{e})$ will become:
\begin{equation}\label{Eqn43}
  \frac{T(\varepsilon_{e})}{T}=1+\frac{(MAv_{s})^{2}TE^{2}}{6mn^{2}e^{6}\bar{z}^{4}\Lambda_{eff}^{2}}
  \left(\frac{\varepsilon_{e}}{T}\right)^{3}=1+B\xi^{3};~
  B=\frac{(MAv_{s})^{2}TE^{2}}{6mn^{2}e^{6}\bar{z}^{4}\Lambda_{eff}^{2}}=
  \frac{MAGT^{2}E^{2}}{6mn^{2}e^{6}\bar{z}^{4}\Lambda_{eff}^{2}};~
  \xi=\frac{\varepsilon_{e}}{T},
\end{equation}
where
\begin{equation}\label{Eqn44}
  f_{0}(\xi)=\left(\exp\biggl(\int_{y}^{\xi}\frac{d\chi}{1+B\chi^{3}}\biggr)+1\right)^{-1}.
\end{equation}
With the account of (\ref{Eqn44}) the normalization condition for the function $f_{0}$, determining also the dependence of the chemical potential $\mu$ from the intensity of electric field ${\bf E}$ becomes:
\begin{equation}\label{Eqn45}
  n_{e}=\bar{z}n=\frac{(2mT)^{3/2}}{2\pi^{2} \hbar^{3}}\int_{0}^{\infty}
  \frac{\xi^{1/2}d\xi}{\exp\biggl(\int_{y}^{\xi}\frac{d\chi}{1+B\chi^{3}}\biggr)+1}.
\end{equation}

Accordingly the final form of the expression (\ref{Eqn38}) for  the electric conduction in a strong electromagnetic field becomes:
\begin{equation}\label{Eqn46}
  \sigma=\frac{T^{3/2}G}{3(2m)^{1/2}\bar{z}e^{2}\Lambda_{eff}(\langle\varepsilon_{e}\rangle)}
  \frac{J'_{3}(y,B)}{J_{1/2}(y,B)},~J_{\nu}(y,B)=\int_{0}^{\infty}\xi^{\nu}f_{0}(\xi,y,B)d\xi;~
  J'_{\nu}(y,B)=\frac{\partial J_{\nu}(y,B)}{\partial y}.
\end{equation}
In (\ref{Eqn46}) the average electron energy $\langle\varepsilon_{e}\rangle$  is determined by the following relation:
\begin{equation}\label{Eqn47}
  \langle\varepsilon_{e}\rangle=T\frac{J_{3/2}(y,B)}{J_{1/2}(y,B)}.
\end{equation}

\section{Discussion}
Let us analyze the limiting cases of the equation set (\ref{Eqn17})-(\ref{Eqn20}) or (\ref{Eqn27})-(\ref{Eqn28}) and (\ref{Eqn19})-(\ref{Eqn20}). In a neutral metal the existence is probable of a two-temperature state when the conduction electrons can have the temperature essentially more than the temperature of the lattice (phonon gas) \cite{Ginzburg:1955,Kaganov:1957}. Such states are observed under interaction of the intensive femto- and picosecond laser radiation with metals. Also in a neutral metal the Maxwell tension tensor of the full electromagnetic field is replaced by the Maxwell tension of the magnetic field  $T_{ik}^{(m)}=(H_{i}H_{k}+\delta_{ik}H^{2}/2)/(4\pi)$ \cite{Landau8:1984}. In the neutral monocrystal metal the phonon energy flow can be neglected in comparison with energy flow of the conduction electrons. However in the case of the polycrystal metal with the grain sizes smaller or equal 100~nm the phonon energy flow inside of the grains cannot be neglected because the mean free path of electrons is equal or more the grain size \cite{Volkov:2010}.

In one-liquid and one-temperature approximation the equation for total energy density will become:
\begin{equation}\label{Eqn29}
  \frac{\partial W^{(\Sigma)}}{\partial t}+{\bf \nabla} {\bf Q}^{(\Sigma)}=
  \frac{m\nu_{ep}}{e^{2}\bar{z}n}j^{2},
\end{equation}
where the total energy flow density in neglect of heat phonon transport becomes:
\begin{equation}\label{Eqn30}
  {\bf Q}^{(\Sigma)}={\bf v}\left(\varepsilon_{\ell}+\bigl<\varepsilon_{p}f_{p}\bigr>+
  \bigl<\varepsilon_{e}f_{e}\bigr>\right)-  \frac{v^{2}}{2}{\bf P}^{(\Sigma)}+
  {\bf v} \bigl({\bf \Pi}^{(\Sigma)}+{\bf T}^{(m)}\bigr)+
  \Bigl<\varepsilon_{e}\frac{\partial \varepsilon_{e}}{\partial {\bf p}}f_{e}\Bigr>.
\end{equation}

At low temperatures when phonon degrees of freedom are frozen the contribution of phonons and the Joule heating to metal energy can be neglected in comparison with elastic energy of a lattice. Then the right part of the equation (\ref {Eqn29})  is equal to zero, and the expressions for ${\bf \Pi}^{(\Sigma)}$ and ${\bf Q}^{(\Sigma)}$ are written in the form, completely in coincidence with corresponding expressions of the nonlinear elasticity  theory of metal advanced in  \cite{Andreev:1985}:
$$\Pi_{ik}^{(\Sigma)}=-T_{ik}^{(m)}+(MA+m\bar{z})nv_{i}v_{k}-\frac{m}{e}\bigl(v_{i}j_{k}+
v_{k}j_{i}\bigr)+2\bigl(\sigma_{\alpha\beta}^{(\ell)}+
\bigl\langle\lambda_{\alpha\beta}^{(e)}f_{e}\bigr\rangle\bigr)a_{i}^{\alpha}a_{k}^{\beta}-
\varepsilon_{\ell}\delta_{ik};$$
$${\bf Q}^{(\Sigma)}={\bf v}\left(\varepsilon_{\ell}+
  \bigl\langle\varepsilon_{e}f_{e}\bigr\rangle\right)-  \frac{v^{2}}{2}{\bf P}^{(\Sigma)}+
  {\bf v} \bigl({\bf \Pi}^{(\Sigma)}+{\bf T}^{(m)}\bigr)+
  \Bigl\langle\varepsilon_{e}\frac{\partial \varepsilon_{e}}{\partial {\bf p}}f_{e}\Bigr\rangle.$$

In the neutral metal it is possible to neglect of the electron inertia. If we also neglect the interband transitions, the generalized Ohm law (\ref{Eqn28}) will become:
\begin{equation}\label{Eqn31}
  {\bf j}=\frac{e^{2}\bar{z}n}{m\nu_{ep}}\left({\bf E}+\frac{1}{c}[{\bf v},{\bf H}]\right)-
  \frac{e}{mc\nu_{ep}}\left[{\bf j},{\bf H}\right]-\frac{e}{m\nu_{ep}}{\bf \nabla} {\bf \Pi}^{e},
\end{equation}
where $\sigma=e^{2}\bar{z}n/(m\nu_{ep})$ is the electric conductivity. The second term in (\ref{Eqn31}) takes into account the contribution of the Hall effect to a current density, and last is the thermoelectricity. In weak magnetic fields the Hall effect can be neglected. If also the gradients of the electronic stresses are small, the Ohm law (\ref{Eqn31}) takes the elementary form known from magnetic hydrodynamics \cite{Landau8:1984}:
\begin{equation}\label{Eqn48}
  {\bf j}=\frac{e^{2}\bar{z}n}{m\nu_{ep}}\left({\bf E}+\frac{1}{c}[{\bf v},{\bf H}]\right)=
  \sigma\left({\bf E}+\frac{1}{c}[{\bf v},{\bf H}]\right).
\end{equation}

In the case of strong pulse electric field and weak magnetic field, the electric conductivity $\sigma$
included in the Ohm law (\ref{Eqn48}) is determined by the formula (\ref{Eqn46}). We shall discuss the behavior of $\sigma$ depending on  ${\bf E}(t)$ value. In weak electric fields ($\mid {\bf E} \mid \ll E_{pF}$) when it is possible to neglect the dependence $\sigma(E^{2})$, the expression (\ref{Eqn48}) coincides with the formula for the electric conductivity obtained in \cite{Volkov:1991,Volkov:2006}:
\begin{equation}\label{Eqn49}
  \sigma=\frac{T^{3/2}G}{3(2m)^{1/2}\bar{z}e^{2}\Lambda_{eff}(\langle\varepsilon_{e}\rangle)}
  \frac{I'_{3}(y)}{I_{1/2}(y)},\quad
  I'_{\nu}(y)=\frac{d I_{\nu}(y)}{d y}.
\end{equation}
In \cite{Volkov:1991} it is shown that the expression (\ref{Eqn49}) has correct asymptotics in the solid metal and the classical ideal plasma, and also describes continuous transition from the condensed state of metal to classical metal plasma. Besides, in \cite{Volkov:2006} it is shown that the values of the electrical conductivity of copper non-ideal plasma calculated with (\ref{Eqn49}) in nonideality parameter range of values $3 \leq \Gamma \leq100$ will be well coordinated with the results of experiments \cite{DeSilva:1994,DeSilva:1998} (see also the report \cite{Redmer:1997} in which these experiments  are analyzed and a comparison with theoretical results of other authors is carried out). Not carrying out detailed quantitative comparison of the expressions (\ref{Eqn48}) and (\ref{Eqn49}) which will be a subject of a separate publication it is possible to draw a conclusion that $\sigma(E^{2})$ will be more or smaller than $\sigma(E^{2}=0)$ only after the vanish of degeneration of the conduction electron gas of the metal (for $|{\bf E}|\geq E_{pF} \sim 1$~MV/cm).

\section{Conclusion}
Thus in the presented work the physical and mathematical model of a metal for the description of fast electrophysical processes taking place under the influence of strong pulse electromagnetic fields was proposed and new wide-range expression for the metal electrical conductivity in a strong electric field being valid both for the quantum plasma and classical one was obtained.

\subsection*{Acknowledgments}
The work carried out within the state order No.~0389-2014-0006 and under the partial financial support of the RFBR (project No.~16-08-00466) and the Ural Branch of RAS within the UB RAS fundamental research program "Matter at high energy densities" (project No. 15-1-2-8).

\section*{References}
\bibliography{NBVolkovs_paper_eng_fin}

\providecommand{\newblock}{}
\begin{thebibliography}{10}
\expandafter\ifx\csname url\endcsname\relax
  \def\url#1{{\tt #1}}\fi
\expandafter\ifx\csname urlprefix\endcsname\relax\def\urlprefix{URL }\fi
\providecommand{\eprint}[2][]{\url{#2}}

\bibitem{Bloembergen:1999}
Bloembergen N 1999 {\em Rev. Mod. Phys.\/} {\bf 71}(2) S283--S287

\bibitem{Mesyats:2005}
Mesyats G~A and Yalandin M~I 2005 {\em Phys. Usp.\/} {\bf 48}(3) 211--232

\bibitem{Milchberg:1988}
Milchberg H~M and et~al 1988 {\em Phys. Rev. Lett.\/} {\bf 61}(20) 2364--2367

\bibitem{Agranat:2007}
Agranat M~B and et~al 2007 {\em JETP Lett.\/} {\bf 85}(6) 271--276

\bibitem{Barakhvostov:2011}
Barakhvostov S~V and et~al 2011 {\em JETP Lett.\/} {\bf 94}(7) 549--555

\bibitem{Volkov:2001}
Volkov N~B 2001 {\em Techn. Phys. Lett.\/} {\bf 27}(3) 236--239

\bibitem{Brandt:2005}
Brandt N~B and Kulbachinskii V~A 2005 {\em Quasiparticles in the Condenced
  Matter Physics\/} (Moscow: Fizmatlit (in Russian))

\bibitem{Andreev:1985}
Andreev A~F and Pushkarov D~I 1985 {\em Sov. Phys. JETP\/} {\bf 62}(5)
  1087--1090

\bibitem{Pushkarov:2001}
Pushkarov D~I 2001 {\em Physics Reports\/} {\bf 354} 411--467

\bibitem{Volkov:2007}
Volkov N~B and et~al 2007 {\em Techn. Phys. Lett.\/} {\bf 33}(1) 69--72

\bibitem{Nezlin:1982}
Nezlin M~V 1982 {\em Beams Dynamics in the Plasma\/} (Moscow: Energoatomizdat
  (in Russian))

\bibitem{Kingsep:1996}
Kingsep A~S 1996 {\em Introduction in the Nonlinear Plasma Physics\/} (Moscow:
  MFTI Publishers (in Russian))

\bibitem{Sedov:1989}
Sedov L~I and Tsypkin A~G 1989 {\em Foundamentals of the Macroscopic Theories
  of Gravitation and Electromagnetism\/} (Moscow: Nauka (in Russian))

\bibitem{Landau:1969}
Landau L~D 1969 {\em Collected Papers, Vol. 1\/} (Moscow: Nauka, pp. 350-383
  (in Russian))

\bibitem{Blokhin:1994}
Blokhin A~M and Dorovskii V~N 1994 {\em Mathematical Simulation Problems in the
  Polyspeed Continuum\/} (Novosibirsk: Institute of Mathematics of SB RAS (in
  Russian))

\bibitem{Volkov:1997}
Volkov N~B 1997 {\em J. Phys. A: Math. Gen.\/} {\bf 30} 6391--6424

\bibitem{Derzhiev:1986}
Derzhiev V~I and et~al 1986 {\em Ion Radiation in a Non-Equilibrium Dense
  Plasma\/} (Moscow: Energoatomizdat (in Russian))

\bibitem{Lifshits:1973}
Lifshits I~M and et~al 1973 {\em Electron Theory of Metals\/} (New York:
  Consultants Bureau)

\bibitem{Landau3:1977}
Landau L~D and Lifshits E~M 1977 {\em Course of Theoretical Physics. Vol. III.
  Quantum Mechanics. Non-relatifistic Theory\/} (Oxford: Pergamon Press)

\bibitem{Landau5:1980}
Landau L~D and Lifshits E~M 1980 {\em Course of Theoretical Physics. Vol. V.
  Statistical Physics, Part 1\/} (Oxford: Pergamon Press)

\bibitem{Klimontovich:1986}
Klimontovich Y~L 1986 {\em Statistical Physics\/} (New York: Harwood Academic)

\bibitem{Braginskii:1965}
Braginskii S~P 1965 {\em Review of Plasma Physics\/} vol~1 ed Leontovich M~A
  (New York: Consultants Bureau) p 205

\bibitem{Landau2:1980}
Landau L~D and Lifshits E~M 1980 {\em Course of Theoretical Physics. Vol. II.
  The Classical Theory of Fields\/} (Amsterdam: Butterworth-Heinemann)

\bibitem{Landau8:1984}
Landau L~D and Lifshits E~M 1984 {\em Course of Theoretical Physics. Vol. VIII.
  Electrodynamics of Continuous Media\/} (Oxford: Pergamon Press)

\bibitem{Shabanskii:1954}
Shabanskii V~P 1954 {\em J. Exptl. Theor. Phys. (U.S.S.R.)\/} {\bf 27} 142

\bibitem{Shabansky:1957}
Shabanskii V~P 1957 {\em Sov. Phys. JETP\/} {\bf 4}(4) 497--508

\bibitem{Ablekov:1980}
Ablekov V~K and et~al 1980 {\em Dokl. Akad. Nauk SSSR\/} {\bf 250}(3) 611--615
  (in Russian)

\bibitem{Ginzburg:1955}
Ginzburg V~L and Shabanskii V~P 1955 {\em Dokl. Akad. Nauk SSSR\/} {\bf 100}
  445--448 (in Russian)

\bibitem{Ginzburg:1970}
Ginzburg V~L 1970 {\em The Propagation of Electromagnetic Waves in Plasmas\/}
  (Oxford: Pergamon Press)

\bibitem{Ginzburg:1955a}
Ginzburg V~L and Silin V~P 1955 {\em JETF\/} {\bf 29} 64--74 (in Russian)

\bibitem{Silin:1956}
Silin V~P 1956 {\em FMM\/} {\bf 3} 193--199 (in Russian)

\bibitem{Silin:1959}
Silin V~P 1959 {\em FMM\/} {\bf 7} 331--334 (in Russian)

\bibitem{Frenkel:1932}
Frenkel J 1932 {\em Wave Mechanics: Elementary Theory\/} (Oxford: Clarendon
  Press)

\bibitem{Harrison:1966}
Harrison W~A 1966 {\em Pseudopotentials in the Theory of Metals\/} (New York:
  Benjamin Press)

\bibitem{Ziman:1972}
Ziman J~M 1972 {\em Principles of the Theory of Solids\/} (Cambridge:
  University Press)

\bibitem{Ziman:1979}
Ziman J~M 1979 {\em Models of Disorder. The Theoretical Physics of
  Homogeneously Disorder Systems\/} (Cambridge: University Press)

\bibitem{Volkov:1979}
Volkov N~B 1979 {\em Techn. Phys.\/} {\bf 49}(9) 2000--2002 (in Russian)

\bibitem{Volkov:1982}
Volkov N~B, Mikhkel'soo V~T and Shneerson G~A 1982 {\em J. Appl. Mech. and
  Techn. Phys.\/} {\bf 23}(5) 607--617

\bibitem{Gerusov:1982}
Gerusov A~V, Ginzburg S~L and Imshennik V~S 1982 {\em Fizika Plasmy\/} {\bf 8}
  487--501 (in Russian)

\bibitem{Lee:1984}
Lee Y~T and More R~M 1984 {\em Phys. Fluids\/} {\bf 27}(5) 1273--1286

\bibitem{Ebeling:1991}
Ebeling W, Feorster A, Fortov V~E, Gryaznov V~K and Polishcuk A~Y 1991 {\em
  Thermophysical Properties of Hot Dense Plasmas\/} (Stutgart, Leipzig:
  Teubner)

\bibitem{Desjarlais:2002}
Desjarlais M~P, Kress J~D and Collins L~A 2002 {\em Phys. Rev. E\/} {\bf 66}
  025401(R)

\bibitem{Mazevet:2005}
Mazevet S and et~al 2002 {\em Phys. Rev. E\/} {\bf 71} 016409

\bibitem{Knyazev:2014}
Knyazev D~V and Levashov P~R 2014 {\em Physics of Plasmas\/} {\bf 21} 073302

\bibitem{Silin:1998}
Silin V~P 1998 {\em An Introduction to Kinetic Theory of Gases\/} (Moscow:
  FIRAN Publishers (in Russian))

\bibitem{Davydov:1937}
Davydov B~I 1937 {\em J. Exptl. Theoret. Phys. (U.S.S.R.)\/} {\bf 7} 1069

\bibitem{Bass:1975}
Bass F~G and Gurevich Y~G 1975 {\em The Hot Electrons and High Electromagnetic
  Waves in a Plasma of Simiconductors and Gas Discharge\/} (Moscow: Nauka (in
  Russian))

\bibitem{Dykman:1981}
Dykman I~M and Tomchuk P~M 1981 {\em Transport Phenomena and Fluctuations in
  Semiconductors\/} (Kiev: Naukova Dumka (in Russian))

\bibitem{Keishiro:1967}
Niu K 1967 {\em Phys. Fluids\/} {\bf 10} 1857--1858

\bibitem{Salat:1969}
Salat A and Kaw P~K 1969 {\em Phys. Fluids\/} {\bf 12} 342--344

\bibitem{Shabansky:1954}
Shabanskii V~P 1954 {\em J. Exptl. Theoret. Phys. (U.S.S.R.)\/} {\bf 27} 147

\bibitem{Volkov:1991}
Volkov N~B and Nemirovsky A~Z 1991 {\em J.Phys. D: Appl. Phys.\/} {\bf 24}
  693--701

\bibitem{Volkov:2006}
Volkov N~B and et~al 2006 {\em Russian Physics Journal\/} {\bf
  49}(11-Suppement) 184--188

\bibitem{Kaganov:1957}
Kaganov M~I and et~al 1957 {\em Sov. Phys. JETP\/} {\bf 4}(2) 173--178

\bibitem{Volkov:2010}
Volkov N~B and et~al 2010 {\em Techn. Phys.\/} {\bf 55}(4) 509--513

\bibitem{DeSilva:1994}
DeSilva A~W and Kunze H~J 1994 {\em Phys. Rev. E\/} {\bf 49}(5) 4448--4454

\bibitem{DeSilva:1998}
DeSilva A~W and Katsouros j~D 1998 {\em Phys. Rev. E\/} {\bf 57}(5) 5945--5951

\bibitem{Redmer:1997}
Redmer R 1997 {\em Physics Reports\/} {\bf 282}(2-3) 35--157

\end{thebibliography}

\end{document}